\documentclass[amsmath,aps,pra,twocolumn]{revtex4-1}
\usepackage{graphicx}
\usepackage{times}
\usepackage{slashed}
\usepackage{bm}

\begin{document}
\title{Radiative corrections to the level width in the presence of magnetic field}

\author{D. Solovyev$^1$, T. Zalialiutdinov$^{1,2}$}
\email[E-mail:]{d.solovyev@spbu.ru}
\affiliation{ 
$^1$ Department of Physics, St. Petersburg State University, Petrodvorets, Oulianovskaya 1, 198504, St. Petersburg, Russia\\
$^2$ Petersburg Nuclear Physics Institute named by B.P. Konstantinov of National Research Centre 'Kurchatov Institut', St. Petersburg, Gatchina 188300, Russia\\
}

\begin{abstract}
We study the influence of constant magnetic field combined with a field induced by the external thermal environment on the atomic decay rates. The importance of radiative corrections, including magnetic interaction, is demonstrated for hydrogen and hydrogen-like ions with low nuclear charge values $Z$. Based on the quantum electrodynamics description, the principal possibility of determining the $g$-factor by observing fluorescence is shown. The considered effects can be used in precision spectroscopic experiments and astrophysical studies.
\end{abstract}

\maketitle
\section{Introduction}

Investigation of radiative decay rates are an inalienable part of precision atomic spectroscopy. Accurate theoretical calculations of transition rates are closely related to measurements of transition frequencies. Experimental efforts aimed at improving the accuracy of lifetime determination in recent years, see, for example, \cite{Trabert_M1,Crespo2005}, make it possible to test quantum electrodynamics (QED) approaches through a detailed comparison of experimental and theoretical results. The experimental uncertainty achieved at the level of one per thousand in few-electron systems \cite{Lapierre} makes such studies sensitive to relativistic, correlation, QED etc. effects \cite{Drag_2003,Tupitsyn2005,Volotka2006,Adkins2008}. The greatest interest in refining the lifetime values of atomic levels manifests itself primarily for metastable states in hydrogen and various highly charged ions. The latter makes it possible to accurately measure transition frequencies between highly excited states \cite{deB-0,deB-1,deB-2,H-exp}.

The importance of the decay rates of a metastable state has been widely demonstrated in the literature. For instance, with the aim of examining the parity non-conservation (PNC) effects, the emission by hydrogen-like atoms in the presence of external electric and magnetic fields has been extensively studied since \cite{Ans,mohr-mix}. Another application is the analysis of the interference of electric and magnetic dipole photons in the presence of an external electric field to measure the Lamb shift in high-Z ions \cite{mohr-mix,HilleyMohr}. Experiments to determine the Lamb shift using the lifetime analysis and the angular distribution of the emitted radiation with respect to the direction of the electric field \cite{drake1977,gould1977} have recently been continued with a significant improvement in accuracy \cite{LambShift-H}. Such precision experiments have as their main goal the determination of fundamental physical constants and, in particular, the determination of the proton charge radius.

Being in the optical region (or close to it), measurements of hyperfine splitting or magnetic dipole transition for fine splitting and the corresponding lifetimes in highly charged ions are also an area of present research (see, for example, \cite{Tupitsyn2005,Lapierre,Volotka2006,Lapierre2006,Mackel2011,Arapoglou}). The precise determination of these quantities opens up possibilities for hydrogen metrology \cite{Weitz,deB-0,deB-1,deB-2,H-exp}, extracting information about the weak interaction from PNC measurements \cite{Gomez2004}, theoretical proposals for finding 'nuclear clock transitions' \cite{Shabaev2022} or probe the QED corrections to the bound-electron $g$-factor \cite{Shabaev1998,Lindenfels2011}.

Separate area of research, where the lifetime of metastable state have a particular importance, can be referred to astrophysical studies of plasma in the early universe \cite{Zeld,Peebles}. No less attention is paid to the two-photon decays of excited states with principal quantum number $n > 2$, whose total contribution to the ionization fraction of primordial plasma reaches the percent level and exceeds the accuracy of the cosmic microwave background measurements \cite{Seager_2000,Chluba2007,Hirata2008,Solovyev2010}. Special attention of researchers is drawn to decay probabilities in atoms placed in external fields. In the astrophysical context, this can primarily be attributed to the study of the matter properties in conditions of huge magnetic fields \cite{Ruder,Wunner_1987} developed into a separate line of research.

In terrestrial laboratory experiments corresponding to precision measurements, control of residual fields is strongly required in order to reduce uncertainties due to magnetic field inhomogeneity \cite{SturmPRA,Sturm2014,Blaum_2020}. Recent spectroscopic experiments aimed at accurately determining fundamental physical constants or testing fundamental interactions necessarily involve a theoretical analysis of level widths \cite{yost,Brandt,Cassidy-Ps}. Accordingly, in this work, we consider the influence of an external magnetic field on the partial transition rates and the lifetimes of excited states in hydrogen-like systems. The description of the lowest order correction to the spontaneous quantity is given within the framework of the rigorous QED theory and includes, in addition to the magnetic field, the field induced by blackbody radiation (BBR).

The paper is organized as follows. In the next section, we discuss the results for the radiative correction of the lowest order in the magnetic field to the spontaneous transition rates. Then, radiative correction involving magnetic interaction for hydrogen-like atoms placed in a thermal equilibrium environment is considered. The conclusions can be found in the last section. Derivations of basic formulas are fetched in the appendices.

\section{Radiative corrections to spontaneous transition rates and level widths in the presence of a magnetic field}

\begin{figure}[hbtp]
\centering
\includegraphics[scale=2]{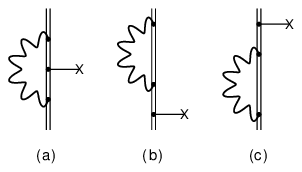}
\caption{Feynman diagrams describing the QED contributions of the order of $\alpha$ to the hyperfine structure splitting and the bound-electron $g$-factor. The imaginary part of these graphs is the radiative correction to the level widths. The tiny line with a cross at the end indicates interaction with an external potential. The doubled solid line denotes the bound electron in the Furry picture. The bold wavy line represents the photon propagator.}
\label{fig1}
\end{figure}

It is well-known that the natural level width of an excited state can be determined as the imaginary part of the one-loop self-energy correction to the energy level of bound electron \cite{LabKlim}. Recently, this approach has been extended to the two-loop level \cite{ZSLP, Jentschura_2002, Jentschura_2008, PhysRevA.79.022510}. Its application to the external thermal environment is considered in \cite{SZATL,ZSL-1ph, ZAS_two-loop-BBR}. On the basis of this formalism, the corresponding radiative QED correction to the level width in the lowest order in the magnetic field is described by the Feynman graphs shown in Fig.~\ref{fig1}.

According to \cite{LabKlim}, the desired correction is determined by the corresponding imaginary part. The details of its derivation can be found in Appendix~\ref{apA} (see Eq. (\ref{23})), and the final result given in the nonrelativistic limit is
\begin{eqnarray}
\label{1.1}
\delta\Gamma_a =4e^2\sum\limits_{n<a}\Delta E_{na}^2
\left|\langle a | \vec{r} | n\rangle\right|^2\left[\langle n| \vec{\mu}\vec{B} |n \rangle
\right.
\\\nonumber
\left.
-\langle a| \vec{\mu}\vec{B} |a \rangle\right],\qquad
\end{eqnarray}
where $\Delta E_{na} \equiv E_n-E_a$, $E_n$ is the nonrelativistic Schr\"{o}dinger energy of the bound electron, $ \langle a | \vec{r} | n\rangle$ is the dipole matrix element, $\vec{B}$ denotes the external magnetic field, $\vec{\mu}$ is the electron magnetic moment. Summation in Eq. (\ref{1.1}) runs over all bound states below an arbitrary excited atomic state $a$. In turn, the partial decay rate corresponds to the terms in the sum in Eq. (\ref{1.1}) with particular value of $n$.

For the homogeneous magnetic field in the nonrelativistic limit the magnetic moment operator is $\vec{\mu}=  -\mu_{\rm B}(\vec{j}+\vec{s})$, where $\mu_{\rm B}$ is the Bohr magneton equal to $1/(2m)$ in relativistic units, $m$ is the electron mass, $ \vec{l} $ and $ \vec{s} $ are the orbital momentum and electron spin operators, $ \vec{j}=\vec{l}+\vec{s} $ is the operator of total angular momentum. Using the results expressed by Eqs. (\ref{13}), (\ref{14}) \cite{VMK} or with the well-known relation \cite{Landau}
\begin{eqnarray}
\label{1.2}
\Delta E_n = \langle n| \vec{\mu}\vec{B} |n \rangle = g_n \mu_{\rm B} B\, m_{j_n},
\end{eqnarray}
it can be found that the summation over the projections included in the sum over $n$ leads to zero. Here $m_{j_n}$ is the projection of the angular momentum $j_n$, $g_n$ denotes the $g$-factor of $n$th state.

With the use of Eq. (\ref{1.2}), however, there are two things to notice. First, a partial transition can be considered at a fixed value of the excited state projection $m_{j_a}$, since Zeeman splitting of atomic levels in the presence of a magnetic field. Second, the Zeeman shift, Eq. (\ref{1.2}), should be taken into account in the $\Delta E_{na}$ factor for both $E_a$ and $E_n$ energies. In the case when the projection $m_{j_a}$ is fixed, there will be linear and quadratic terms in the field. Conversely, when summing over all projections $m_{j_a}$, $m_{j_n}$, only the contribution that is quadratic in the field survives.

Considering as an example the Ly$_\alpha$ transition ($2p-1s$ decay in hydrogen), one can find for $j_a=1/2$ and $j=3/2$
\begin{eqnarray}
\label{1.3}
\frac{\delta\Gamma_{2p_{1/2}}}{\mu_{\rm B} B^2} &=& \frac{2^{14}e^2}{3^{10}}\left(3 g_{2p_{1/2}}^2 + 2 g_{2p_{1/2}} g_{1s_{1/2}} + 3 g_{1s_{1/2}}^2\right),\qquad
\\
\nonumber
\frac{\delta\Gamma_{2p_{3/2}}}{\mu_{\rm B} B^2} &=& \frac{2^{15}e^2}{3^{10}}\left(15 g_{2p_{3/2}}^2 - 10 g_{2p_{3/2}} g_{1s_{1/2}} + 3 g_{1s_{1/2}}^2\right),
\end{eqnarray}
respectively. Here we have used $ E_{2p_{1/2}} - E_{1s_{1/2}} = 3/8$ in the lowest order and the radial part of the matrix element $\langle a | \vec{r} | n\rangle$ gives $2^7/3^4\sqrt{2/3}$ in atomic units (see appendices ~\ref{apA} and \ref{apB} for the details).

The expression (\ref{1.3}) illustrates how lifetime measurements can be used to determine $g$-factors for various excited states. When $g_{1s_{1/2}}$, the experimental level widths, and the magnetic field are known, the equations (\ref{1.3}) can be solved with respect to $g_{2p_{1/2}}$, $g_{2p_{3/2}}$. The most problematic in this case is that lifetimes are experimentally determined with less accuracy than transition frequencies. However, by observing the behavior of the level width in the field, one can find the left side of equations (\ref{1.3}).

Below, we present the results of numerical calculations of the partial transition rates and the level widths for $2p_{1/2}$ and $2p_{3/2}$ excited states of a hydrogen atom and several hydrogen-like ions. The results are presented in Table~\ref{tab:1} for magnetic fields $1$ G ($10^{-4}$ T) and $100$ G ($10^{-2}$ T), partial transition rates for a fixed projection of the initial states, $m_{j_a}$, and level widths (summed over all projections) are given in s$^{-1}$.
\begin{table}
\caption{Numerical values of the radiative correction Eq. (\ref{1.1}) for the Ly$_{\alpha}$ transition rate in the hydrogen atom and some hydrogen-like ions ($Z=2,\,6$). In the first column, the transition under consideration is given by the designation of the total angular momenta for the initial and final states ($2p_j\rightarrow 1s_{1/2}$). Each picked out line contains the radiative correction summed over all projections, and the next lines give partial values for a fixed initial projection $m_{j_a}$. The projection values are shown in the second column, and the partial transition rates are given in fourth and fifth columns for a magnetic field of $1$ G and $100$ G, respectively.}
\label{tab:1}
\begin{tabular}{c c c c c }
\hline
Ly$_\alpha$ & $m_{j_a}$ & $\Gamma_a^{\rm nat}$, s$^{-1}$ & $\delta\Gamma^{1 {\rm G}}_a$, s$^{-1}$ & $\delta\Gamma^{100 {\rm G}}_a$, s$^{-1}$  \\
\hline
\hline
 &  &   $ Z=1 $ &  &  \\
 \hline
 \hline 
 $1/2\rightarrow 1/2$ &  & $6.2681\times 10^8$ & $-0.5690$ & $-5689.8$ \\
\hline
 & $1/2$ & -- & $-1.3354\times 10^4$ & $-1.3411\times 10^6$ \\
 & $-1/2$ & -- & $1.3353\times 10^4$ & $1.3297\times 10^6$ \\
 \hline
 \hline
$3/2\rightarrow 1/2$ &  & $6.2681\times 10^8$ & $-0.4267$ & $-4267.5$ \\
\hline
 & $3/2$ & -- & $-2.0031\times 10^4$ & $-2.0074\times 10^6$ \\
 & $1/2$ & -- & $-6677.4$ & $-6.7197\times 10^5$ \\
 & $-1/2$ & -- & $6676.6$ & $6.6344\times 10^5$ \\
 & $-3/2$ & -- & $2.0031\times 10^4$ & $1.9988\times 10^6$ \\
 
\hline
\hline
 &  &   $ Z=2 $ &  &  \\
 \hline
 \hline 
 $1/2\rightarrow 1/2$ &  & $1.0029\times 10^{10}$ & $-0.5689$ & $-5689.2$ \\
\hline
 & $1/2$ & -- & $-5.3413\times 10^4$ & $5.3469\times 10^6$ \\
 & $-1/2$ & -- & $5.3412\times 10^4$ & $5.3355\times 10^6$ \\
 \hline
 \hline
$3/2\rightarrow 1/2$ &  & $1.0029\times 10^{10}$ & $-0.4267$ & $-4267.3$ \\
\hline
 & $3/2$ & -- & $-8.0126\times 10^4$ & $-8.0168\times 10^6$ \\
 & $1/2$ & -- & $-2.6709\times 10^4$ & $-2.6751\times 10^6$ \\
 & $-1/2$ & -- & $2.6708\times 10^4$ & $2.6666\times 10^6$ \\
 & $-3/2$ & -- & $8.0125\times 10^4$ & $8.0083\times 10^6$ \\ 
  
\hline
\hline
 &  &   $ Z=6 $ &  &  \\
 \hline
 \hline 
 $1/2\rightarrow 1/2$ &  & $8.1237\times 10^{11}$ & $-0.5683$ & $-5683.0$ \\
\hline
 & $1/2$ & -- & $-4.8047\times 10^5$ & $-4.8053\times 10^7$ \\
 & $-1/2$ & -- & $4.8047\times 10^5$ & $4.8041\times 10^7$ \\
 \hline
 \hline
$3/2\rightarrow 1/2$ &  & $8.1237\times 10^{11}$ & $-0.4266$ & $-4266.3$ \\
\hline
 & $3/2$ & -- & $-7.2129\times 10^5$ & $-7.2134\times 10^7$ \\
 & $1/2$ & -- & $-2.4043\times 10^5$ & $-2.4047\times 10^7$ \\
 & $-1/2$ & -- & $2.4043\times 10^5$ & $2.4039\times 10^7$ \\
 & $-3/2$ & -- & $7.2129\times 10^5$ & $7.2125\times 10^7$ \\ 
\hline
\end{tabular}
\end{table}

The physical meaning of $\delta\Gamma_a$, Eq. (\ref{1.1}), can be understood as the interference of photons emitted by the unperturbed and Zeeman states of the atom. Thus, there is nothing surprising in the fact that this quantity can be negative. The total contribution $\delta\Gamma_a$ can be obtained from the partial transition rates by averaging over the projection $m_{j_a}$, i.e. summing the values listed in Table~\ref{tab:1} with the coefficient $1/(2j_a+1)$. It can also be noted that the natural level width in a purely nonrelativistic representation (the sum of $nl\rightarrow n'l'$ transitions) can be obtained using the sum rule \cite{Sob,Solovyev2011}. The natural partial widths corresponding to different values of the total angular momentum are equal in the cases under consideration. 

Table~\ref{tab:1} shows the results for the hydrogen atom and hydrogen-like ions with low $Z$ values. This is primarily due to the use of the nonrelativistic approximation to derive the formula (\ref{1.1}), which allows us to employ the Schr\"{o}dinger spectrum and wave functions. Thus, the accuracy of $\delta\Gamma_a$ calculations is limited by relativistic corrections of the order of $(Z\alpha)^2$ (radiative QED corrections are also omitted). 

\begin{figure}[hbtp]
\centering
\includegraphics[width=0.9\columnwidth]{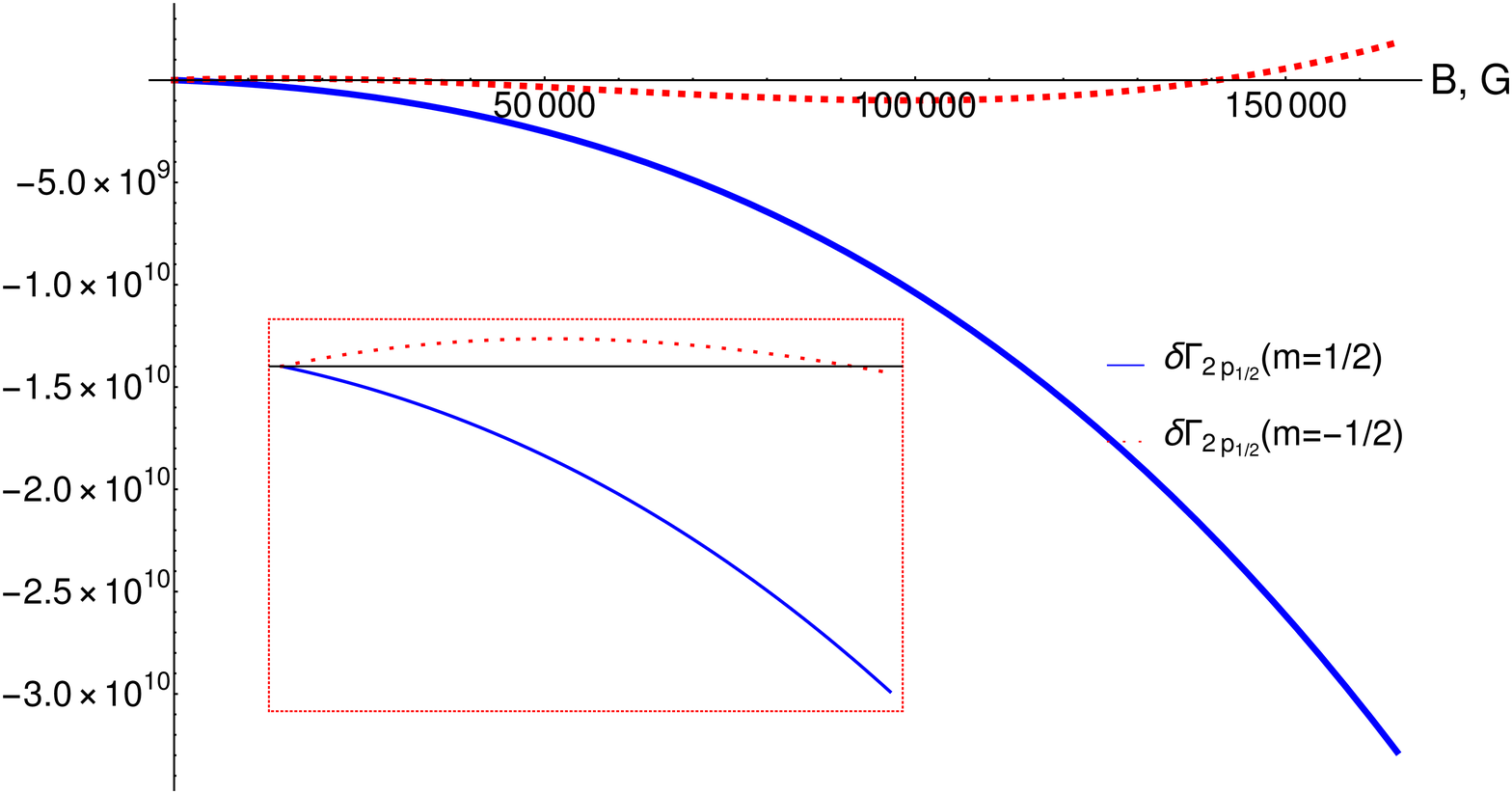}
\caption{Behavior of partial $2p_{1/2}\rightarrow 1s_{1/2}$ transition rates in a magnetic field for the hydrogen atom. The solid line corresponds to the projection of the total angular momentum $m_{j_a}=1/2$, and the dotted line corresponds to $m_{j_a}=-1/2$. For clarity, the selected frame shows an enlarged area close to zero.}
\label{fig2}
\end{figure}

\begin{figure}[hbtp]
\centering
\includegraphics[width=0.9\columnwidth]{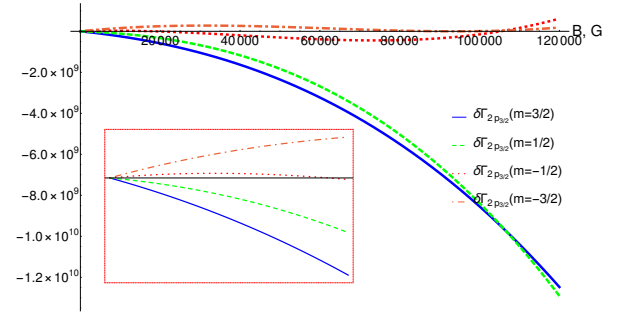}
\caption{Behavior of partial $2p_{3/2}\rightarrow 1s_{1/2}$ transition rates in a magnetic field for the hydrogen atom. The solid line corresponds to the projection of the total angular momentum $m_{j_a}=3/2$, the dashed line - $m_{j_a}=1/2$, the dotted line corresponds to $m_{j_a}=-1/2$, and the dotted line with dots - $m_{j_a}=-3/2$. For clarity, the selected frame shows an enlarged area close to zero.}
\label{fig3}
\end{figure}

As follows from Table~\ref{tab:1}, the obtained values of the radiative correction to the total level width $\delta\Gamma_a$ are almost independent of $Z$. It can be found with the parametric estimation for Eq. (\ref{1.1}): $\Delta E_{na}\sim m(\alpha Z)^2$ and $\langle r \rangle\sim1/(m\alpha Z)$ in relativistic units (here $\alpha$ is the fine structure constant and $m$ is the electron mass). Then, summing over projections and taking into account the Zeeman shift of energies $E_n$, $E_a$ (see formula (\ref{1.2})), the factor $\Delta E_{na}^2$ is replaced by $\Delta E_{na} \mu_{\rm B}B$, which reduces the dependence on $Z$ by $\langle r \rangle^2$ in the leading order. In turn, for partial transition rates, this dependence is expressed by the factor $Z^2$ with the same estimates (there is no reduction in this case). Finally, as expected, the radiative correction (\ref{1.1}) to the total level width is quadratic in the field and almost linear for partial quantities. This behavior persists until $\approx 10^4$ G ($1$ T), when the quadratic contribution to the partial probabilities starts to outweigh the linear one, see Fig.~\ref{fig2} for the $2p_{1/2}$ state and Fig.~\ref{fig3} for the $2p_{3/2}$ state in the hydrogen atom. 

Note that perturbation theory for the hydrogen atom is valid up to magnitude of matrix element  $\langle n| \vec{\mu}\vec{B} |n \rangle\approx E_n-E_a\sim 1/2$ (in atomic units). The latter condition corresponds to the maximum magnetic field $B\approx 2\times 10^9$ G ($\mu_{\rm B} = 2.12716\times 10^{-6}$ a.u./T). However, a more stringent constraint follows from the fact that the correction Eq. (\ref{1.1}) (given that it is negative) should not exceed the value of the natural level width. Equating the absolute values of the expression (\ref{1.1}) and the natural level width, the maximum allowable field strength is $B\approx 3.3$ T ($33190.9$ G) for the $2p_{1/2}$ state.

\section{Radiative corrections to spontaneous transition rates and level widths in the presence of a magnetic field combined with the field induced by the blackbody radiation}
\label{B_BBR}

Taking into account the astrophysical interest in the level widths and decay rates in atoms, this section considers the thermally stimulated radiative correction. The corresponding quantities can be obtained using the technique of Feynman diagrams and adiabatic S-matrix approach. The analytical derivations for the imaginary part of these diagrams representing $\delta\Gamma_a^\beta$ can follow the calculations given in Appendix~\ref{apA}. The only difference is that in the case of finite temperature QED for bound states the photon propagator, depicted by the wavy line in Fig.~\ref{fig1}, should be replaced by a thermal one \cite{DonH}. Details of derivations and description of formalism can be found in Appendix~\ref{apB}. 

The final expression for the finite temperature radiative correction to the level width in the presence of an additional homogeneous magnetic field is (see Eq. (\ref{31}))
\begin{eqnarray}
\label{2.1}
\delta\Gamma^\beta_a = \frac{4e^2}{3}\sum\limits_n \left[\langle n| \vec{\mu}\vec{B} |n \rangle-\langle a| \vec{\mu}\vec{B} |a \rangle\right]|\langle a|\vec{r}|n\rangle|^2
\times\qquad
\\
\nonumber
\left[ \beta \Delta E_{na}^3 n^2_\beta(|\Delta E_{na}|) e^{\beta\, |\Delta E_{na}|} - 3 \Delta E_{na}^2 n_\beta(|\Delta E_{na}|) \right],
\end{eqnarray}
where $n_{\beta}(\omega)=(\mathrm{exp}(\beta\omega)-1)^{-1}$ is the photon density number of BBR field (Planck distribution function), $\beta=1/k_{\mathrm{B}}T$, $k_{\mathrm{B}}$ is the Boltzmann constant in relativistic units, $T$ is the radiation temperature in kelvin. In contrast to Eq. (\ref{1.1}) the summation in Eq.  (\ref{2.1}) extends to all states (above and below an arbitrary state $a$ under consideration), including the continuum. This immediately leads to the conclusion that the ground state has a BBR stimulated level width due to a possible excitation process, see \cite{Jent-Stark}.

Assuming astrophysical applications of the correction Eq. (\ref{2.1}), the numerical results for $\delta\Gamma^\beta_a$ are collected in Table~\ref{tab:2} for various low-lying states in hydrogen and hydrogen-like ions at different temperatures and magnetic field $100$ G.
\begin{widetext}
\begin{center}
\begin{table}
\caption{Numerical values of the radiative correction Eq. (\ref{2.1}) for various low-lying states in hydrogen and hydrogen-like ions ($Z=2,\,6$) at different temperatures and a magnetic field of $100$ G. The first column shows the atomic states for which the correction (\ref{2.1}) is calculated. The following columns show the values of correction $\delta\Gamma_a^\beta$ at a fixed temperature. Each first substring contains the values of the level widths $\Gamma_{a}^{\beta} $ stimulated by blackbody radiation (see \cite{Farley}), the second substring contains the values of Eq. (\ref{2.1}). The natural level width is given below All values are given in s$^{-1}$.}
\label{tab:2}
\begin{tabular}{l l l l l}
\hline
$\Gamma_a^{\rm nat}$ in s$^{-1}$ & $\Gamma^\beta_a$ in s$^{-1}$ at $300$ K & at $500$ K & at $1000$ K & at $3000$ K  \\
State & $\delta\Gamma^\beta_a$ in s$^{-1}$  &   &   &    \\
\hline
\hline
  &  & $Z=1$ ($100$ G) & &  \\
 \hline
 \hline 
 $\Gamma^{\rm nat}_{1s}=0$ &  $\Gamma^\beta_{1s}\sim 10^{-163}$ & $\sim 10^{-94}$ & $7.04\times 10^{-43}$ & $1.35\times 10^{-8}$ \\
 $1s_{1/2}$ & $\delta\Gamma^\beta_{1s}\sim 10^{-170}$ & $\sim 10^{-100}$ & $-6.29\times 10^{-49}$ & $-1.21\times 10^{-15}$ \\
     
\hline
 $\Gamma^{\rm nat}_{2s}=8.229$ & $\Gamma^\beta_{2s}=1.420\times 10^{-5}$ & $2.367\times 10^{-5}$ & $0.0202$ & $4.701\times 10^4$ \\
 $2s_{1/2}$ & $\delta\Gamma^\beta_{2s}=-0.0104$ & $-0.0173$ & $-0.0346$ & $-0.1052$ \\
 
\hline
$\Gamma^{\rm nat}_{2p}=6.268\times 10^8$ & $\Gamma^\beta_{2p}=4.734\times 10^{-6}$ & $7.891\times 10^{-6}$ & $0.0329$ & $7.583\times 10^4$ \\
$2p_{1/2}$ & $\delta\Gamma^\beta_{2p_{1/2}}=-0.0041$ & $-0.0069$ & $-0.0138$ & $-0.0425$ \\

$2p_{3/2}$ & $\delta\Gamma^\beta_{2p_{3/2}}=-0.0031$ & $-0.0052$ & $-0.0104$ & $-0.0329$ \\

 \hline
 \hline
  &  & $Z=2$ ($100$ G) & &  \\
 \hline
 \hline 
 
 $\Gamma^{\rm nat}_{1s}=0$ & $\Gamma^\beta_{1s}=0$ & $0$ & $\sim 10^{-196}$ & $8.147\times 10^{-59}$ \\
   $1s_{1/2}$   &  $\delta\Gamma^\beta_{1s}=0$ & $0$ & $\sim 10^{-204}$ & $-8.199\times 10^{-66}$ \\
     
\hline
 $\Gamma^{\rm nat}_{2s}=526.7$ & $\Gamma^\beta_{2s}=6.252\times 10^{-4}$ & $1.042\times 10^{-3}$ & $2.085\times 10^{-3}$ & $6.474\times 10^{-3}$ \\
$2s_{1/2}$ & $\delta\Gamma^\beta_{2s}=-0.0026$ & $-0.0043$ & $-0.0086$ & $-0.0259$ \\
 
\hline  
$\Gamma^{\rm nat}_{2p}=1.003\times 10^{10}$ & $\Gamma^\beta_{2p}=2.084\times 10^{-4}$ & $3.475\times 10^{-4}$ & $6.952\times 10^{-4}$ & $2.439\times 10^{-3}$  \\
$2p_{1/2}$ & $\delta\Gamma^\beta_{2p_{1/2}}=-0.0010$ & $-0.0017$ & $-0.0034$ & $-0.0104$ \\
 
$2p_{3/2}$ & $\delta\Gamma^\beta_{2p_{3/2}}=-0.0008$ & $-0.0013$ & $-0.0026$ & $-0.0078$ \\

 \hline
 \hline
  &  & $Z=6$ ($100$ G) & &  \\
 \hline
 \hline 
 
     
 $\Gamma^{\rm nat}_{2s}=3.839\times 10^5$ & $\Gamma^\beta_{2s}=2.032\times 10^{-1}$ & $3.475\times 10^{-1}$ & $7.082\times 10^{-1}$ & $2.152$ \\
$2s_{1/2}$ & $\delta\Gamma^\beta_{2s}=-0.0005$ & $-0.0008$ & $-0.0018$ & $-0.0057$ \\
 
\hline  
$\Gamma^{\rm nat}_{2p}=8.124\times 10^{11}$ & $\Gamma^\beta_{2p}=6.774\times 10^{-2}$ & $1.158\times 10^{-1}$ & $0.236\times 10^{-1}$ & $7.172\times 10^{-1}$  \\
$2p_{1/2}$ & $\delta\Gamma^\beta_{2p_{1/2}}=-0.0001$ & $-0.0002$ & $-0.0004$ & $-0.0011$ \\
 
$2p_{3/2}$ & $\delta\Gamma^\beta_{2p_{3/2}}=-5.6\times 10^{-5}$ & $-0.0001$ & $-0.0003$ & $-0.0008$ \\  
  
\hline
\end{tabular}
\end{table}
\end{center}
\end{widetext}

First, the values listed in Table~\ref{tab:2} show the behavior of the correction Eq. (\ref{2.1}) depending on the nuclear charge $Z$. It can be found that $\delta\Gamma^\beta_{a}$ decreases with increasing $Z$, becoming negligible for highly charged hydrogen-like ions. This is due to the decrease in the Planck distribution function $n_\beta$ as $\Delta E_{na}\sim Z^2$ increases. The values of $\delta\Gamma^\beta_{a}$ grow slowly with increasing temperature. However, the correction Eq. (\ref{2.1}) is significant for $Z=1,\,2$ and larger than the widths induced by BBR. Thus, one can expect its significance for describing the recombination epoch in the early Universe \citep{Seager_2000}. In this case, the role of the external field $\vec{B}$ is played by the primary magnetic fields. It is known that they are present in almost all types of cosmic structures in the Universe - from small planets to galaxies and the largest clusters of galaxies. There is reason to believe that even at early stages the warmed Universe is permeated with such fields, albeit of a smaller magnitude \cite{PhysRevLett.116.191302}.  Their origin and mechanism of generation in the epoch of matter recombination has been widely discussed in the literature \cite{zeldovich1983fluid, parker2019cosmical, Sethi, Gopal}.

Finally, the width of the ground state becomes noticeable at $3000$ K for a hydrogen atom and is infinitely small for any other value of $Z$ at any temperature. Note also that the values given in Table~\ref{tab:2} correspond to a magnetic field of $100$ G, which is rather weak. Rough estimates for other field strengths can be obtained taking into account the quadratic behavior in $B$, which is feasible approximately up to $10^4$ G (see the previous section).

To complete the analysis, consider the correction given by Eq. (\ref{2.1}) as a function of $B$ at a fixed temperature. This can be attributed to laboratory experiments, where the influence of the external environment can be ascribed to blackbody radiation at room temperature. Here we also carry out calculations for partial quantities, i.e. with a fixed projection of the initial state angular momentum, $m_{j_a}$. Numerical values are given in Table~\ref{tab:3} for low-lying states in hydrogen and hydrogen-like ions with $Z=2,\,6$.
\begin{widetext}
\begin{center}
\begin{table}
\caption{Numerical values of the radiative correction Eq. (\ref{2.1}) in the hydrogen atom and some hydrogen-like ions ($Z=2,\,6$) depending on the magnetic field $B$ at room temperature. The first column gives the initial state with a fixed projection, $m_{j_a}$. The following columns show the partial widths for field strengths given in gauss. All values are given in s$^{-1}$.}
\label{tab:3}
\begin{tabular}{c c c c c c }
\hline
State & $1$ G & $10$ G & $100$ G & $10^3$ G & $10^4$ G  \\
\hline
\hline
 &  & &  $Z=1$ ($300$ K) &  &  \\
 \hline
 \hline 
 $2s_{1/2}^{m_{j_a}=1/2}$ & $2.081\times 10^{-4}$ & $1.987\times 10^{-3}$ & $1.054\times 10^{-2}$ & $-1.937\times 10^{-1}$ & $34.612$ \\
 
 $2s_{1/2}^{m_{j_a}=-1/2}$ & $-2.101\times 10^{-4}$ & $2.195\times 10^{-3}$ & $-3.128\times 10^{-2}$ & $-1.094$ & $-48.152$ \\
\hline

 $2p_{1/2}^{m_{j_a}=1/2}$ & $1.564\times 10^{-4}$ & $1.527\times 10^{-3}$ & $1.153\times 10^{-2}$ & $6.138\times 10^{-2}$ & $29.596$ \\
 
 $2p_{1/2}^{m_{j_a}=-1/2}$ & $-1.572\times 10^{-4}$ & $-1.610\times 10^{-3}$ & $-1.983\times 10^{-2}$ & $-5.716\times 10^{-1}$ & $-34.803$ \\
 \hline

 $2p_{3/2}^{m_{j_a}=3/2}$ & $2.349\times 10^{-4}$ & $2.321\times 10^{-3}$ & $2.041\times 10^{-2}$ & $7.598\times 10^{-2}$ & $28.727$ \\
 
 $2p_{3/2}^{m_{j_a}=1/2}$ & $7.810\times 10^{-5}$ & $7.530\times 10^{-4}$ & $4.729\times 10^{-3}$ & $8.214\times 10^{-2}$ & $24.612$ \\
 
 $2p_{3/2}^{m_{j_a}=-1/2}$ & $-7.872\times 10^{-5}$ & $-8.152\times 10^{-4}$ & $-3.895\times 10^{-1}$ & $6.138\times 10^{-2}$ & $-28.256$ \\
 
 $2p_{3/2}^{m_{j_a}=-3/2}$ & $-2.355\times 10^{-4}$ & $-2.383\times 10^{-3}$ & $-2.663\times 10^{-2}$ & $-5.463\times 10^{-1}$ & $-33.417$ \\
 
\hline
\hline
 &  &  & $Z=2$ ($300$ K) &  &  \\
 \hline
 \hline 
 $2s_{1/2}^{m_{j_a}=1/2}$ & $6.926\times 10^{-4}$ & $6.903\times 10^{-3}$ & $6.670\times 10^{-2}$ & $4.344\times 10^{-1}$ & $-8.654$ \\
 
 $2s_{1/2}^{m_{j_a}=-1/2}$ & $-6.932\times 10^{-4}$ & $-6.955\times 10^{-3}$ & $-7.187\times 10^{-2}$ & $-9.514\times 10^{-1}$ & $-32.749$ \\
 \hline

 $2p_{1/2}^{m_{j_a}=1/2}$ & $5.196\times 10^{-4}$ & $5.186\times 10^{-3}$ & $5.093\times 10^{-2}$ & $4.163\times 10^{-1}$ & $-5.997\times 10^{-1}$ \\
 
 $2p_{1/2}^{m_{j_a}=-1/2}$ & $-5.198\times 10^{-4}$ & $-5.207\times 10^{-3}$ & $-5.300\times 10^{-2}$ & $-6.231\times 10^{-1}$ & $-15.517$ \\
 \hline

$2p_{3/2}^{m_{j_a}=3/2}$ & $7.794\times 10^{-4}$ & $7.787\times 10^{-3}$ & $7.718\times 10^{-2}$ & $7.019\times 10^{-1}$ & $2.714\times 10^{-2}$ \\
 
 $2p_{3/2}^{m_{j_a}=1/2}$ & $2.598\times 10^{-4}$ & $2.591\times 10^{-3}$ & $2.521\times 10^{-2}$ & $1.823\times 10^{-1}$ & $5.573\times 10^{-1}$ \\
 
 $2p_{3/2}^{m_{j_a}=-1/2}$ & $-2.599\times 10^{-4}$ & $-2.606\times 10^{-3}$ & $-2.676\times 10^{-2}$ & $-3.374\times 10^{-1}$ & $-10.333$ \\
 
 $2p_{3/2}^{m_{j_a}=-3/2}$ & $-7.796\times 10^{-4}$ & $-7.803\times 10^{-3}$ & $-7.873\times 10^{-2}$ & $-8.570\times 10^{-1}$ & $-15.537$ \\
\hline
\hline
 &  &  & $Z=6$ ($300$ K) &  &  \\
 \hline
 \hline 
 $2s_{1/2}^{m_{j_a}=1/2}$ & $1.40751\times 10^{-1}$ & $1.407$ & $14.067$ & $1.399\times 10^2$ & $1.322\times 10^3$ \\
 
 $2s_{1/2}^{m_{j_a}=-1/2}$ & $-1.40752\times 10^{-1}$ & $-1.408$ & $-14.084$ & $-1.416\times 10^2$ & $-1.492\times 10^3$ \\
 \hline

 $2p_{1/2}^{m_{j_a}=1/2}$ & $2.93232\times 10^{-3}$ & $2.9322\times 10^{-2}$ & $2.931\times 10^{-1}$ & $2.923$ & $28.376$ \\
 
 $2p_{1/2}^{m_{j_a}=-1/2}$ & $-2.93234\times 10^{-3}$ & $-2.9324\times 10^{-2}$ & $-2.933\times 10^{-1}$ & $-30.266$ & $-15.517$ \\
 \hline

$2p_{3/2}^{m_{j_a}=3/2}$ & $4.39849\times 10^{-3}$ & $4.39843\times 10^{-2}$ & $4.3978\times 10^{-1}$ & $4.391$ & $43.275$ \\
 
 $2p_{3/2}^{m_{j_a}=1/2}$ & $1.46616\times 10^{-3}$ & $1.46609\times 10^{-2}$ & $1.4655\times 10^{-1}$ & $1.459$ & $13.951$ \\
 
 $2p_{3/2}^{m_{j_a}=-1/2}$ & $-1.46617\times 10^{-3}$ & $-1.46624\times 10^{-2}$ & $-1.4669\times 10^{-1}$ & $-1.473$ & $-15.368$ \\
 
 $2p_{3/2}^{m_{j_a}=-3/2}$ & $-4.39850\times 10^{-4}$ & $-4.39857\times 10^{-2}$ & $-4.3992\times 10^{-1}$ & $-4.406$ & $-44.692$ \\
\hline
\hline
  
\end{tabular}
\end{table}
\end{center}
\end{widetext}

First, it follows from Table~\ref{tab:3} that the correction $\delta\Gamma^\beta_a$ for a fixed projection $m_{j_a}$ is significant even at room temperature and a weak magnetic field for the metastable state $2s$ in hydrogen and hydrogen-like ions with low $Z$. The value of $\delta\Gamma^\beta_a(m_{j_a})$ should be compared with the natural level width and the decay rate $\Gamma_a^\beta$ stimulated by BBR \cite{Farley}, see Table~\ref{tab:2}. For example, in a hydrogen atom in a magnetic field $\sim 1$ G at room temperature, the correction Eq. (\ref{2.1}) for the upper projection is an order of magnitude greater than $\Gamma_{2s}^\beta$ (same the most for the lower projection, but with the opposite sign). The values of $\delta\Gamma^\beta_{2s}(m_{j_{2s}})$ and $\Gamma_{2s}^\beta$ become comparable at $Z=2,\,6$, which can be explained the different parameterization by $Z$ for these quantities. Thus, one can expect the contribution (\ref{2.1}) to be insignificant for large $Z$. 

The radiative correction Eq. (\ref{2.1}) can also be compared with the leading order correction to the two-photon decay rate of the $2s$ state in the hydrogenlike atomic systems \cite{Jent-E1E1}. It can be found that the results shown in Tables~\ref{tab:2} and \ref{tab:3} are significantly higher than the radiative logarithmic correction to the level width analyzed in \cite{Jent-E1E1}. In turn, using the example of a helium ion ($Z=2$), one can note that the values for $\delta\Gamma^\beta_{2s}$ and $\delta\Gamma^\beta_{2s}(m_{j_{2s}})$ found in this work significantly exceed the upper limit imposed on the amplitude of parity non-conserving $2p$ admixture \cite{Hinds}. At room temperature and a magnetic field of $1$ G, the correction $\delta\Gamma^\beta_{2s}$ is comparable to $|\delta|< 2.4\times 10^{-5}$ \cite{Hinds} and increases quadratically with field strength $B$ (for partial quantities, linear growth takes place).

The values listed in Table~\ref{tab:2} can be obtained from Table~\ref{tab:3} by summing the respective partial transition rates, which is the sum over the state projection $a$. In view of the strong numerical reduction in Table~\ref{tab:3}, where necessary, more significant digits are given. The accuracy of our calculations can be estimated as relativistic corrections of the leading order, the relative value of which for $\delta\Gamma^\beta_a$ can be roughly given by $(\alpha Z)^2$. The latter is determined by the nonrelativistic limit used to derive the basic formulas. 

\begin{figure}[hbtp]
\centering
\includegraphics[scale=0.3]{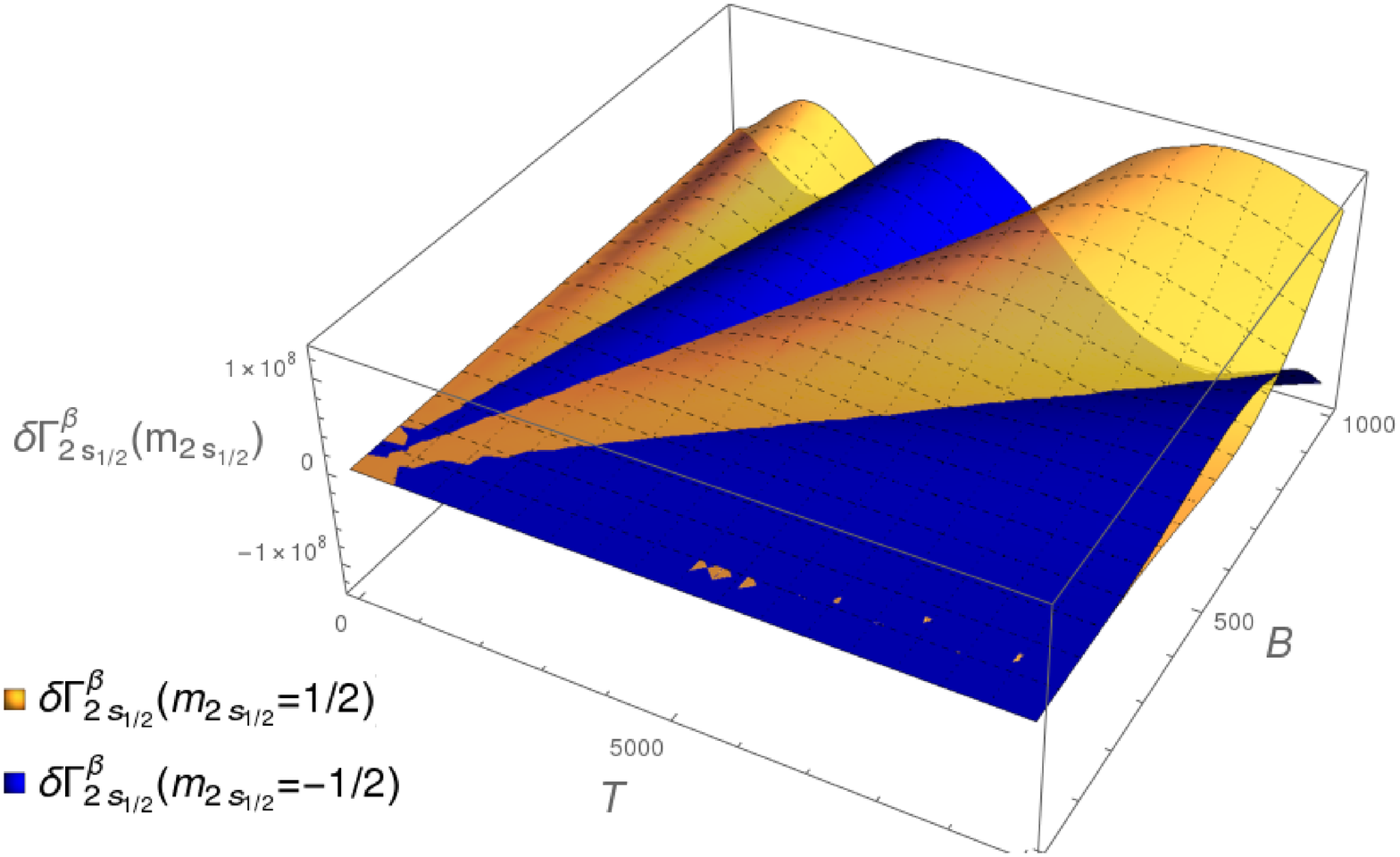}
\caption{Behavior of the correction $\delta\Gamma^\beta_{2s_{1/2}}$ to the level width at a fixed projection $m_{j_{2s_{1/2}}}$ in a magnetic field at different temperatures in the hydrogen atom. The orange (online colored) surface (started from zero to the upper half-space) corresponds to the positive value of the projection, and the blue (going to the lower half-space near zero) represents $\delta\Gamma^\beta_{2s_{1/2}}$ for the negative projection.}
\label{fig4}
\end{figure}

Finally, for a visual demonstration of the dependence on the field and temperature, partial corrections (\ref{2.1}) for the $2s$ state at fixed projections $m_{j_{2s_{1/2}}}=1/2$ and $m_{j_{2s_{1/2}}}=-1/2$ are shown in Fig.~\ref{fig4}.
In particular, the value corresponding to the positive projection at low temperatures and field strengths is in the upper half-space and vice versa for the negative projection. As the magnetic field strength increases, the quadratic and cubic contributions (the Zeeman shift Eq. (\ref{1.2}) should be taken into account) become significant, leading to the 'waves' shown in Fig.~\ref{fig4}. The amplitude of the 'wave' also increases with temperature. The intersection $\delta\Gamma^\beta_{2s}(m_{j_{2s}}=1/2)$ and $\delta\Gamma^\beta_{2s}(m_{j_{2s}}=-1/2)$ shows that there are field values screening the temperature, making the contribution Eq. (\ref{2.1}) equal to zero.

\section{Conclusions}

In this paper, we considered radiative corrections to the partial and total level widths in the presence of a magnetic field for hydrogen and a hydrogen-like atom placed in a thermal environment. The radiative corrections can be represented by the Feynman graphs shown in Fig.~\ref{fig1}, where the vertex part illustrates the magnetic interaction, and the photon loop can be attributed to blackbody radiation at zero and finite temperatures.

Analyzing first the case of zero temperature, using the Ly$_{\alpha}$ decay rate as an example, the significance of the correction for $Z=1,\,2,\,6$ was demonstrated. The values given in Table~\ref{tab:1} show that even at a weak field strength of $100$ G, the partial quantity is two orders of magnitude smaller than the natural transition probability, thus representing a weighty value of line broadening. The correction Eq. (\ref{1.1}) for the total level width can be estimated as a function linear in $Z$, and the partial transition probability is proportional to $Z^2$. This leads to the diminishing role of $\delta\Gamma_a$ with increasing nuclear charge for hydrogen-like atomic systems, see also Table~\ref{tab:1}. The dependence of the decay rates of the $2p_{1/2}$ and $2p_{3/2}$ states with a fixed projection in hydrogen on the magnetic field strength is illustrated in Figs.~\ref{fig2} and \ref{fig3}.

\begin{figure}[hbtp]
\centering
\includegraphics[scale=0.2]{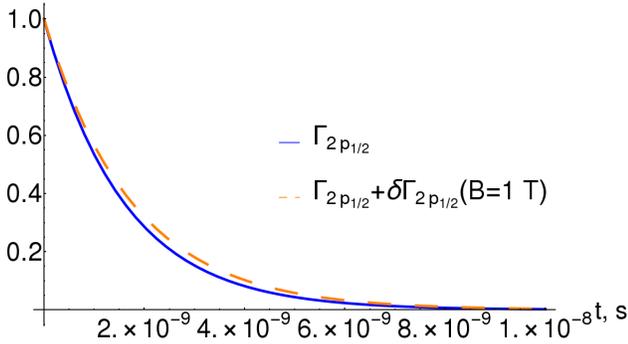}
\caption{Lyman-alpha time decay of the $2p_{1/2}$ state in a hydrogen atom at a magnetic field stregth $1$ T. The solid (blue online) line represents the decay according to the natural level width, the dotted (orange online) line corresponds to the sum of the natural level width and the correction Eq. (\ref{1.1}). The values of time are given in seconds and the value $\Gamma_{2p_{1/2}} = 6.2681\times 10^8$ s$^{-1}$ is used.}
\label{fig5}
\end{figure}

\begin{figure}[hbtp]
\centering
\includegraphics[scale=0.2]{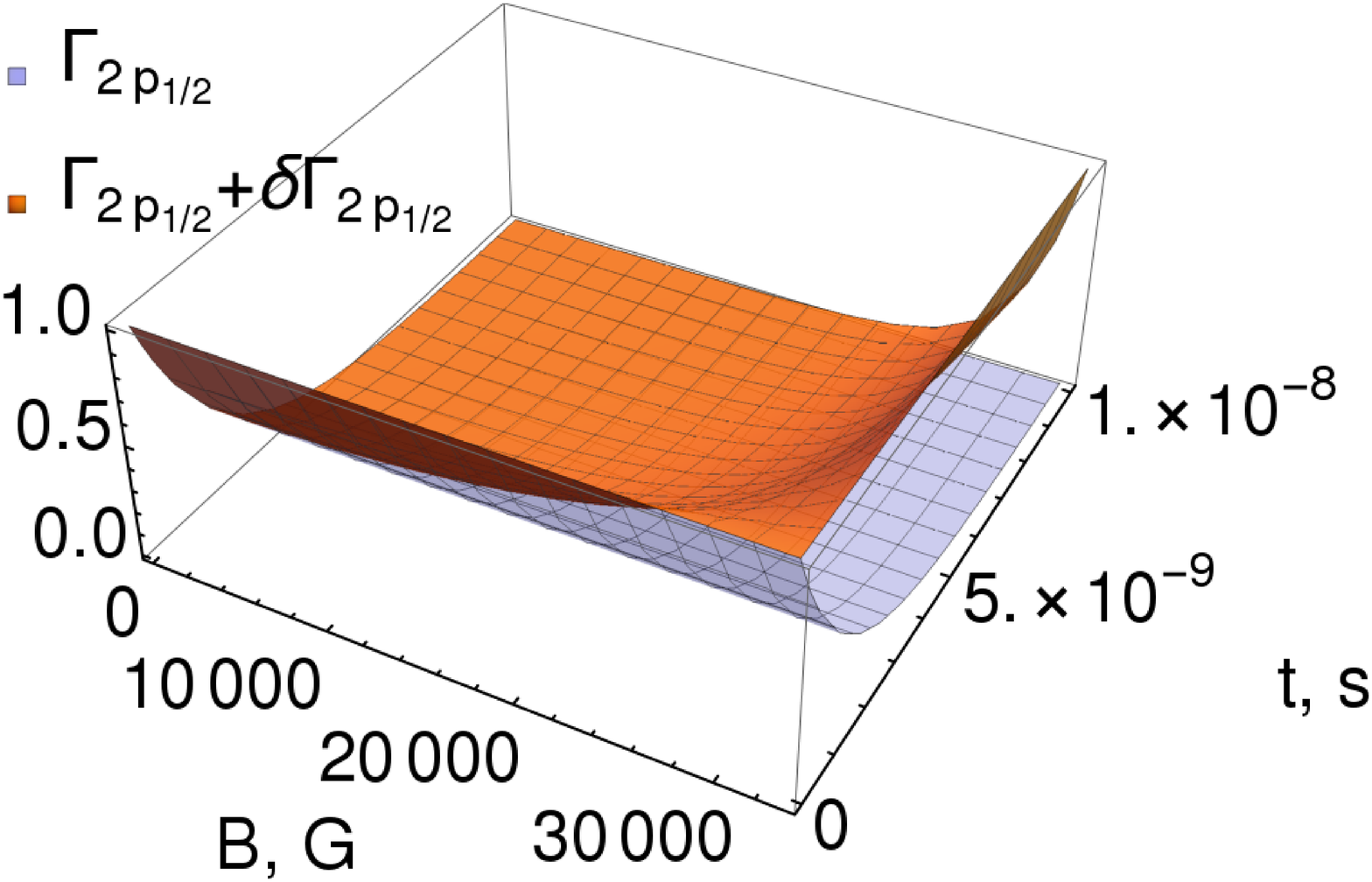}
\caption{Lyman-alpha time decay of the $2p_{1/2}$ state in a hydrogen atom as a function of magnetic field and time. The designations are the same as in Fig.~\ref{fig5}, the magnetic field is given in gauss. The bottom (blue online) surface represents the natural decay, and the top (orange online) includes the correction Eq. (\ref{1.1}).}
\label{fig6}
\end{figure}

However, the most interesting result is that the correction (\ref{1.1}) opens up the possibility of determining the $g$-factor in experiments of a completely different type, i.e. by measuring the fluorescence of excited states of an atom in a magnetic field. Taking into account the Ly$_\alpha$ transition in hydrogen, the theoretical part for such measurements can be expressed by Eq. (\ref{1.3}). It can be assumed that there are no significant difficulties for accurate theoretical calculations (including relativistic and QED estimates) of the quantities included in Eq. (\ref{1.3}). The most problematic in this proposal is the experimental accuracy of the decay rate measurement. The effect can be detected by observing the decay time dependence in the absence and deviations in the presence of a magnetic field. For example, the time dependence of the decay of the $2p_{1/2}$ state in the hydrogen atom is shown in Fig.~\ref{fig5} at a magnetic field strength of $B=1$ T,  while the graph showing the decay as a function of the magnetic field and time is illustrated in Fig.~\ref{fig6}.

\begin{figure}[hbtp]
\centering
\includegraphics[scale=0.15]{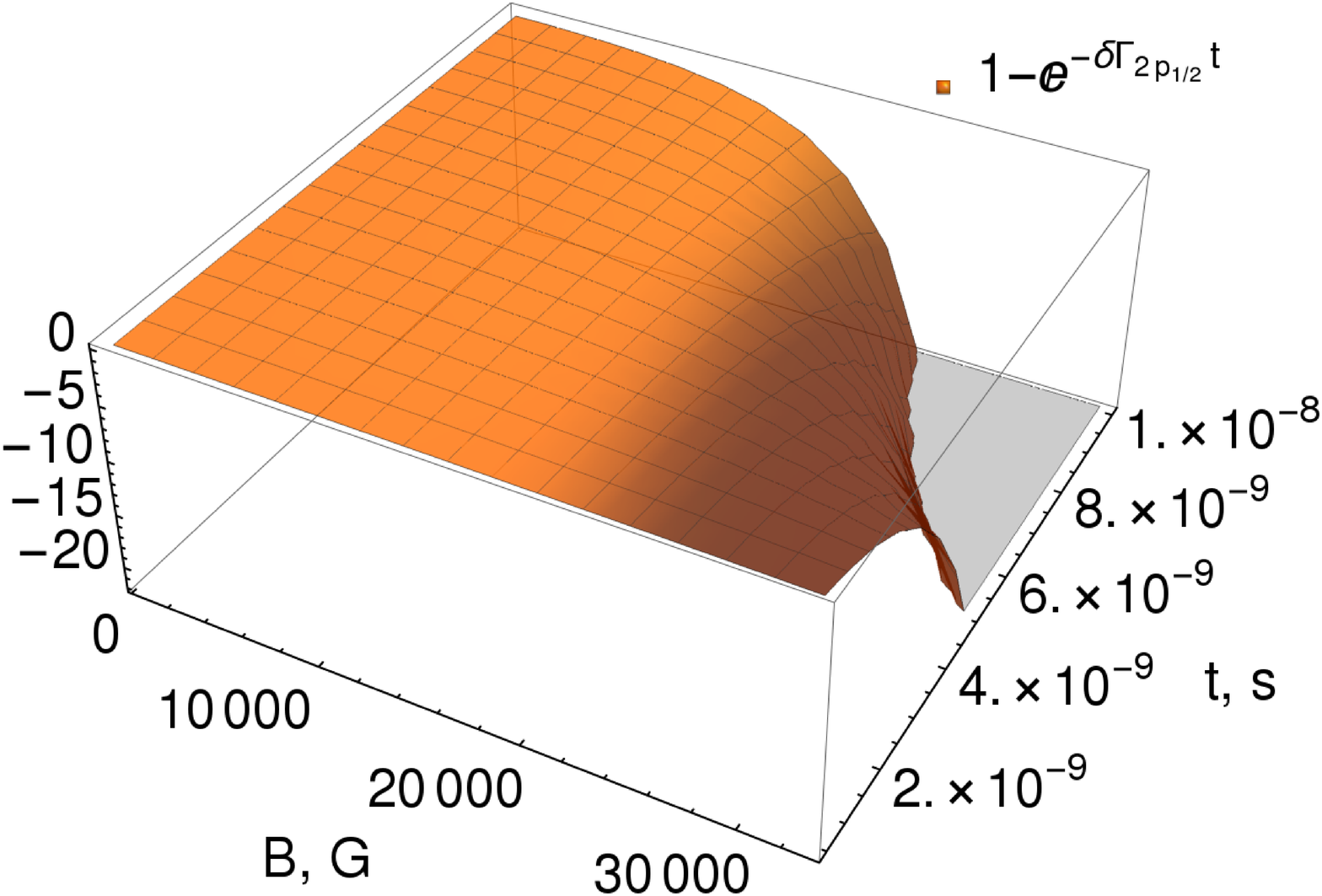}
\caption{Lyman-alpha time decay of the $2p_{1/2}$ state in a hydrogen atom as a function of magnetic field and time. The designations are the same as in Fig.~\ref{fig5}, the magnetic field is given in Gauss. The bottom (blue online) surface represents the natural decay, and the top (orange online) includes the correction Eq. (\ref{1.1}).}
\label{fig7}
\end{figure}

Compiling the relative difference between natural decay and decay in an external magnetic field $\delta=\left(e^{-\Gamma_a t}-e^{-\Gamma_a t+\delta\Gamma_a(B) t}\right)/e^{-\Gamma_a t} = 1- e^{-\delta\Gamma_a(B) t}$, an exponentially increasing difference (in absolute value) should be observed with time and increasing field strength. For clarity, Fig.~\ref{fig7} shows the relative difference between the natural and corrected decays of the $2p_{1/2}$ excited state in the hydrogen atom as a function of time and magnetic field.

Finally, in section~\ref{B_BBR} a detailed analysis of the magnetic field influence on hydrogen and hydrogen-like ions placed in a thermal environment is given. The latter is important for several reasons. Most importantly, experiments with atomic systems are hard to shield from external microwave radiation at room temperature. According to the results given in Tables~\ref{tab:2}, \ref{tab:3}, the contribution of the total radiative correction $\delta\Gamma^\beta_{2s}$ in hydrogen is about $10^{-2}$ (at $T=300$ K and $B=100$ G), decreasing with $Z$ and increasing with temperature and a magnetic field strength. The dependence of partial quantities $\delta\Gamma^\beta_{2s_{1/2}}(m_{2s_{1/2}})$ in the hydrogen atom is shown in Fig.~\ref{fig4}. In addition, it can be noted that the analysis performed in this work can be used in experiments aimed at accurately studying the effects of parity non-conservation \cite{Ans,Hinds} and astrophysical problems, where magnetic fields and temperatures can exceed laboratory ones by orders of magnitude.

{\it Acknowledgements.} This work was supported by the Russian Science Foundation under grant No. 22-12-00043.

\appendix
\renewcommand{\theequation}{A\arabic{equation}}
\setcounter{equation}{0}

\bibliography{mybibfile}

\section{Basic formulas for the radiative correction to the level width in the presence of a magnetic field}
\label{apA}

One of the ways to determine the radiative QED correction to the level width of an arbitrary state $a$ in the presence of a magnetic field corresponds to considering the imaginary part of the graphs shown in Fig.~\ref{fig1}.
\begin{figure}[hbtp]
\centering
\includegraphics[scale=2]{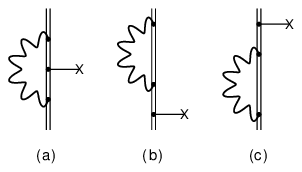}
\caption{Feynman diagrams describing the QED contributions of the order of $\alpha$ to the hyperfine structure splitting and the bound-electron $g$-factor. The imaginary part of these graphs is the radiative correction to the level widths. The tiny line with cross at the end indicates interaction with an external potential. The doubled solid line denotes the bound electron in the Furry picture. The bold wavy line represents the photon propagator.}
\label{fig1a}
\end{figure}

Following the adiabatic S-matrix approach the third order correction to the energy level of bound state $a$ is \cite{Gell,Low,Sucher,g-ZGS}
\begin{eqnarray}
\label{1}
\Delta E_a^{(3)} = 3\langle \Phi_a^{0}| \hat{S}_{\eta}^{(3)}|\Phi_a^{0}\rangle
\\
\nonumber
 -
3\langle\Phi_a^{0}| \hat{S}_{\eta}^{(2)}|\Phi_a^{0}\rangle
\langle\Phi_a^{0}| \hat{S}_{\eta}^{(1)}|\Phi_a^{0}\rangle 
+\langle\Phi_a^{0}| \hat{S}_{\eta}^{(1)}|\Phi_a^{0}\rangle^3,
\end{eqnarray}
where $\Phi_a^{0}$ is the solution of the unperturbed Hamiltonian, $\hat{S}_{\eta}$ is the evolution operator  
\begin{eqnarray}
\label{2}
\hat{S}_{\eta}=\mathrm{T}\left[\mathrm{exp}\left(-\mathrm{i}e\int d^4x e^{-\eta |t|}\hat{H}_{\mathrm{i}}(x)  \right) \right]
\\
\nonumber
=
1+\sum\limits_{k=1}^{\infty}\frac{(-\mathrm{i}e)^k}{k!}\int d^4x_{k}\dots \int d^4x_{1} 
\\
\nonumber
\times
e^{-\eta |t_{k}|}\dots e^{-\eta |t_{1}|}\mathrm{T}\left[\hat{H}_{\mathrm{i}}(x_{k})\dots \hat{H}_{\mathrm{i}}(x_{1})\right]
.
\end{eqnarray}
Here $\mathrm{T}[\dots]$ denotes the time-ordered product of interaction density $\hat{H}_{\mathrm{i}}$, which is
\begin{eqnarray}
\label{3}
\hat{H}_{\mathrm{i}}(x)=\hat{j}^{\mu}(x)(\hat{A}_{\mu}(x)+\hat{A}_{\mu}^{\mathrm{pert}}(x)).
\end{eqnarray}

In Eq. (\ref{3}) $\hat{A}_{\mu}(x)$ and $\hat{A}_{\mu}^{\mathrm{pert}}(x))$ denote the operators of photon field and external perturbation, respectively. The operator of electron current in Eq. (\ref{3}) is defined as follows
\begin{eqnarray}
\label{4}
\hat{j}^{\mu}(x)=-\frac{1}{2}e[\hat{\overline{\psi}}(x)\gamma^{\mu},\hat{\psi}(x)]
,
\end{eqnarray}
where $\hat{\overline{\psi}}(x)=\hat{\psi}^{\dagger}(x)\gamma_{0}$, $\hat{\psi}(x)$ is the operator of fermion field, where $\psi$ and $\overline{\psi}$ are the one-electron and its Dirac conjugated wave functions, $\gamma^{\mu}$ is the Dirac gamma matrices and $x=(t,\vec{r})$ represents the four-space coordinate vector.

Below, we deal with only one-electron atomic systems and, therefore, the graphs with more than one fermionic line in the initial and final states are excluded. In addition, we keep only the terms of the first order in the perturbing potential $\hat{A}_{\mu}^{\mathrm{pert}}$. 

Matrix elements up to $\hat{S}_{\eta}^{(3)}$ in different orders in $e$ are equal to the next expressions:
\begin{eqnarray}
\label{5}
\langle \Phi_a^{0}| \hat{S}_{\eta}^{(1)}|\Phi_a^{0}\rangle
 = -\mathrm{i}e\int d^4x\,\overline{\psi}_{a}(x) e^{-\eta|t|} \gamma^{\mu} A_{\mu}^{\mathrm{pert}}(x) \psi_a(x),\qquad
\end{eqnarray}
represents contribution of the first order. The second-order contribution (self-energy correction):
\begin{eqnarray}
\label{6}
\langle \Phi_a^{0}| \hat{S}_{\eta}^{(2)}|\Phi_a^{0}\rangle
 = (-\mathrm{i}e)^2\int d^4x_{1}d^4x_{2}\overline{\psi}_{a}(x_{1})\, e^{-\eta|t_{1}|}\times
\\
\nonumber
\gamma^{\mu_1}D_{\mu_{1}\mu_{2}}(x_{1},x_{2})e^{-\eta|t_{2}|}\gamma^{\mu_{2}}S(x_{1},x_{2})
e^{-\eta|t_{2}|} \psi_a(x_{2}).
\end{eqnarray}
The third-order matrix elements are given by
\begin{eqnarray}
\label{7}
\langle \Phi_a^{0}| \hat{S}_{\eta}^{(3)}|\Phi_a^{0}\rangle_{\mathrm{Fig.3(a)}}
 = (-\mathrm{i}e)^3\int d^4x_{1}d^4x_{2}d^4x_{3}\times 
\\
\nonumber
\overline{\psi}_{a}(x_{1}) e^{-\eta|t_{1}|}
\gamma^{\mu_1}D^{\beta}_{\mu_{1}\mu_{3}}(x_{1},x_{3})e^{-\eta|t_{3}|}\gamma^{\mu_{3}}\times
\\
\nonumber
S(x_{1},x_{2}) e^{-\eta|t_{2}|} \gamma^{\mu_{2}}A_{\mu_{2}}^{\mathrm{pert}}(x_2)S(x_{2},x_{3})\psi_a(x_{3})
\end{eqnarray}
and
\begin{eqnarray}
\label{8}
\langle \Phi_a^{0}| \hat{S}_{\eta}^{(3)}|\Phi_a^{0}\rangle_{\mathrm{Fig.3(b)}} 
= 
\langle \Phi_a^{0}| \hat{S}_{\eta}^{(3)}|\Phi_a^{0}\rangle_{\mathrm{Fig.3(c)}}
\\
\nonumber
=
(-\mathrm{i}e)^3\int d^4x_{1}d^4x_{2}d^4x_{3}\times
\\
\nonumber
\overline{\psi}_{a}(x_{1}) e^{-\eta|t_{1}|} \gamma^{\mu_1}D^{\beta}_{\mu_{1}\mu_{3}}(x_{1},x_{2}) e^{-\eta|t_{3}|} \gamma^{\mu_{3}}\times
\\
\nonumber
S(x_{1},x_{2}) e^{-\eta|t_{2}|} \gamma^{\mu_{3}}A_{\mu_{3}}^{\mathrm{pert}}(x_2)S(x_{2},x_{3})\psi_a(x_{3}).
\end{eqnarray}

Here $\psi_{a}(x)=\psi(\vec{r})_{a} e^{\mathrm{i}E_{a}t} $, $ \psi(\vec{r})_{a}$ is a solution of the Dirac equation for an one-electron ion with charge $Z$ in a state with energy $E_{a}$. The electron propagator $S(x_1,x_2)$ is 
\begin{eqnarray}
\label{9}
S(x_{1}x_{2})=\frac{\mathrm{i}}{2\pi}\int\limits_{-\infty}^{+\infty}d\Omega\, e^{-\mathrm{i}\Omega(t_{1}-t_{2})}
\sum\limits_{n}\frac{\psi_{n}(\vec{r}_1)\overline{\psi}_{n}(\vec{r}_2)}{\Omega-E_{n}(1-\mathrm{i}0)}
,\qquad
\end{eqnarray}
and the sum over $n$ in Eq. (\ref{9}) runs over all the entire Dirac spectrum. The photon propagator in coordinate space representation (in Feynman gauge, see \cite{LabKlim}), can be reduced to
\begin{eqnarray}
\label{10}
D_{\mu\nu}(x_1 x_2)=\frac{g_{\mu\nu}}{2\pi\mathrm{i} r_{12}}\int\limits_{-\infty}^{\infty}d\omega\, 
e^{\mathrm{i} |\omega|r_{12}-\mathrm{i}\omega(t_{1}-t_{2})}
,\qquad
\end{eqnarray}
where $r_{12}\equiv |\vec{r}_1-\vec{r}_2|$.

Then, substituting Eqs. (\ref{5})-(\ref{8}) into Eq. (\ref{1}), the total energy shift can be written as a sum of vertex (Fig.~\ref{fig1a} (a)) and wave-function (Fig.~\ref{fig1a} (b), (c)) contributions \cite{lindgren}. Performing integration over time variables and taking the limit $\eta \rightarrow 0 +$ \cite{LSP-sep}, the expression for the vertex type correction reads
\begin{eqnarray}
\label{11}
\Delta E_{a}^{\mathrm{ver}} = \frac{e^3 }{2\pi\mathrm{i}}
\sum\limits_{n,m}\int\limits_{-\infty}^{\infty}
d\omega 
\frac{\langle a m| \frac{1-\vec{\alpha}_1\vec{\alpha}_2}{r_{12}} e^{\mathrm{i}|\omega| r_{12}} |n a\rangle}{E_a-\omega-E_n(1-\mathrm{i}0)}
\\
\nonumber
\times
\frac{\langle n|\gamma^{\nu}A_{\nu}^{\mathrm{pert}}|m \rangle}{E_a-\omega-E_m(1-\mathrm{i}0)},
\end{eqnarray}
where $\vec{\alpha}$ denotes the vector of Dirac matrices. Similar evaluation for the wave-function part yields
\begin{eqnarray}
\label{12}
\Delta E_{a}^{\mathrm{wf}}
= \frac{e^3 }{\pi\mathrm{i}} \int\limits_{-\infty}^{\infty}
d\omega \sum\limits_{\stackrel{n, m}{m\neq a}}
\frac{\langle a n| \frac{1-\vec{\alpha}_1\vec{\alpha}_2}{r_{12}} e^{\mathrm{i}|\omega| r_{12}} |n m\rangle}{E_{a}-\omega-E_{n}(1-\mathrm{i}0)}
\nonumber
\\
\times
\frac{\langle m| \gamma^{\nu}A_{\nu}^{\mathrm{pert}} |a \rangle}{E_{a}-E_{m}}\qquad
\\
\nonumber
- \frac{e^3 }{2\pi\mathrm{i}} \int\limits_{-\infty}^{\infty}
d\omega \sum\limits_n 
\frac{\langle an| \frac{1-\vec{\alpha}_1\vec{\alpha}_2}{r_{12}} e^{\mathrm{i}|\omega| r_{12}} |na\rangle
\langle a| \gamma^{\nu}A_{\nu}^{\mathrm{pert}} |a \rangle
}{[E_{a}-\omega-E_{n}(1-\mathrm{i}0)]^2}, \qquad
\end{eqnarray}
where, the 'reference-state' contribution, $m = a$, is presented by the second term in Eq. (\ref{12}), see \cite{Blunden} for details.

For a specific interaction, $e\gamma^{\nu}A_{\nu}^{\mathrm{pert}} = \vec{\mu}\vec{B}$, where $\vec{B}$ is the homogeneous magnetic field, and the magnetic moment operator $\vec{\mu}$ in the nonrelativistic limit is given by $\vec{\mu}=-\mu_{\rm B}(\vec{l}+2\vec{s}) = -\mu_{\rm B}(\vec{j}+\vec{s})$ (here $\mu_{\rm B}$ is the Bohr magneton equal to $1/(2m)$ in relativistic units, $m$ is the electron mass, $ \vec{l} $ and $ \vec{s} $ are the orbital momentum and electron spin operators, $ \vec{j}=\vec{l}+\vec{s} $ is the operator of total angular momentum), one can obtain \cite{VMK}:
\begin{eqnarray}
\label{13}
\langle n'l's'j'm_{j'}|j_{z}|nlsjm_{j}\rangle = 
\\
\nonumber
\delta_{n'n}\delta_{l'l}\delta_{s's}\delta_{j'j}
\sqrt{\frac{j(j+1)(2j+1)}{2j'+1}} C_{jm\, 10}^{j'm_{j'}}\, ,
\end{eqnarray}
\begin{eqnarray}
\label{14}
\langle n'l's'j'm_{j'}|s_{z}|nlsjm_{j}\rangle = 
\delta_{n'n}\delta_{l'l}\delta_{s's}
\\
\nonumber
\times
(-1)^{2j'+l+s+1-m_{j'}}
\Pi_{j'}\Pi_{j}\sqrt{s(s+1)(2s+1)}
\\\nonumber
\times
\begin{pmatrix}
j'          & 1 & j     \\
-m_{j'}     & 0 & m_{j}
\end{pmatrix}
\begin{Bmatrix}
s      & l & j     \\
j'     & 1 & s
\end{Bmatrix}
.
\end{eqnarray}
The above expressions use standard notations: $\Pi_{a\,b\dots}=\sqrt{(2a+1)(2b+1)\dots}$, $C_{j_1m_1\,j_2m_2}^{j_3m_3}$ is the Clebsch-Gordan coefficient for the decomposition of $|jm\rangle$ in terms of $|j_1m_1\rangle |j_2 m_2\rangle$; in Eq. (\ref{14}), the factor given in parentheses refers to the ordinary $3j$-symbol, and the curly braces denote the $6j$-symbol \cite{VMK}. 

According to Eqs. (\ref{13}), (\ref{14}), only diagonal matrix elements survive for the $\vec{\mu}\vec{B}$ interaction in Eqs. (\ref{11}), (\ref{12}). Thus, the first term in Eq. (\ref{12}) vanishes, and the expression (\ref{11}) reduces to the second term in Eq. (\ref{12}) with opposite sign and interaction averaged over states other than $a$, i.e.
\begin{eqnarray}
\label{15}
\Delta E_a = \Delta E_{a}^{\mathrm{ver}} + \Delta E_{a}^{\mathrm{wf}} =\qquad
\\
\nonumber
\frac{e^2 }{2\pi\mathrm{i}} \int\limits_{-\infty}^{\infty}
d\omega \sum\limits_n 
\frac{\langle an| \frac{1-\vec{\alpha}_1\vec{\alpha}_2}{r_{12}} e^{\mathrm{i}|\omega| r_{12}} |na\rangle}{[E_{a}-\omega-E_{n}(1-\mathrm{i}0)]^2}
\times
\\
\nonumber
\left[\langle n| \vec{\mu}\vec{B} |n \rangle-\langle a| \vec{\mu}\vec{B} |a \rangle\right].
\end{eqnarray}
To evaluate further expression (\ref{15}) can be simplified with the relation:
\begin{eqnarray}
\label{16}
\frac{1}{[E_{a}-\omega-E_{n}(1-\mathrm{i}0)]^2}\qquad
\\\nonumber
= -\frac{\partial}{\partial E_a}\frac{1}{E_{a}-\omega-E_{n}(1-\mathrm{i}0)}.\qquad
\end{eqnarray}
Then, the integration over poles can be performed in the similar way as in \cite{LabKlim}.  

Denoting $I_{na}(r_{12}) = \int\limits_{-\infty}^\infty d\omega\,e^{\mathrm{i}|\omega| r_{12}}/\left(E_{n}(1-\mathrm{i}0)-E_{a}+\omega\right)$, one can find
\begin{eqnarray}
\label{17}
I_{na}(r_{12}) = \frac{\mathrm{i}\pi}{2}\left(1+\frac{E_n}{|E_n|}\right)\left(1-\frac{\Delta E_{na}}{|\Delta E_{na}|}\right)e^{\mathrm{i}x} 
\\
\nonumber
+
2\mathrm{i}\frac{\Delta E_{na}}{|\Delta E_{na}|}\left[\mathrm{ci}\left(x\right)\sin\left(x\right)-\mathrm{si}\left(x\right)\cos\left(x\right)\right],
\end{eqnarray}
where $x\equiv |\Delta E_{na}|r_{12}$, $\Delta E_{na}\equiv E_n-E_a$ and the sine and cosine integrals are defined as
\begin{eqnarray}
\label{18}
\mathrm{si}\left(x\right) = -\int\limits_x^\infty \frac{\sin t}{t}dt,\qquad
\mathrm{ci}\left(x\right) = -\int\limits_x^\infty \frac{\cos t}{t}dt.
\end{eqnarray}
In the nonrelativistic approximation, when $\Delta E_{na}r_{12}\approx \alpha Z\ll 1$ (for $E_n>0$ and neglecting the negative energy spectrum), the expansion into the series of $I_{na}(r_{12})$ gives
\begin{eqnarray}
\label{19}
I_{na}(r_{12})\approx \pi\mathrm{i} -2\mathrm{i}\Delta E_{na}r_{12}- \pi\left(|\Delta E_{na}|-\Delta E_{na}\right)r_{12} \qquad
\nonumber
\\
+
 2\mathrm{i}\Delta E_{na}r_{12}\ln\left(\gamma|\Delta E_{na}|r_{12}\right) - \frac{\pi\mathrm{i}}{2}\left(\Delta E_{na}r_{12}\right)^2 \qquad\qquad
\\
+
\nonumber
 \frac{\mathrm{i}}{2}\left(\Delta E_{na}r_{12}\right)^3 + \frac{\pi}{6}\left(|\Delta E_{na}|-\Delta E_{na}\right)\Delta E_{na}^2r_{12}^3 \qquad
\\
\nonumber
-
\frac{\mathrm{i}}{3}\left(\Delta E_{na}r_{12}\right)^3\ln\left(\gamma|\Delta E_{na}|r_{12}\right),\qquad
\end{eqnarray}
where $\gamma=0.57721\dots$ is the Euler's constant. 

The radiative correction to an arbitrary atomic state $a$ occurs when the imaginary part of (\ref{19}) is substituted into the expression (\ref{15}). The resulting real contribution diverges and requires a renormalization procedure \cite{shabaev-report}. In turn, the real part of $I_{na}(r_{12})$ converges and leads to the imaginary part of the energy shift, which can be related to the radiative correction to the excited state level width. 

The imaginary part ($\mathrm{Im}\Delta E_a$) in the nonrelativistic limit arises as
\begin{eqnarray}
\label{20}
\mathrm{Im}\Delta E_a = -\frac{e^2}{2}\sum\limits_n \left[\langle n| \vec{\mu}\vec{B} |n \rangle-\langle a| \vec{\mu}\vec{B} |a \rangle\right]\qquad
\\
\nonumber
\times
\frac{\partial}{\partial E_a}\left(|\Delta E_{na}|-\Delta E_{na}\right)\qquad
\\\nonumber\times
\langle a n | \left(1-\vec{\alpha}_1\vec{\alpha}_2-\frac{\beta^2_{na}r_{12}^2}{6}\right)|n a\rangle,
\end{eqnarray}
where we have neglected the product of the $\vec{\alpha}$-matrix and the last term in parentheses. Then, using the relations $\Delta E_{na}\left(\vec{r}\right)_{an} = \mathrm{i}\left(\vec{p}\right)_{an}$, $r_{12}^2=r_1^2+r_2^2-2\vec{r}_1\vec{r}_2$ and the orthogonality property of wave functions, it can be found
\begin{eqnarray}
\label{21}
\mathrm{Im}\Delta E_a =\frac{e^2}{3}\frac{\partial}{\partial E_a}\sum\limits_n\left(|\Delta E_{na}|-\Delta E_{na}\right)\Delta E_{na}^2
\\
\nonumber
\times
\langle a | \vec{r}_1 | n\rangle\langle n | \vec{r}_2 | a\rangle\left[\langle n| \vec{\mu}\vec{B} |n \rangle-\langle a| \vec{\mu}\vec{B} |a \rangle\right].
\end{eqnarray}

Finally, with $\sum\limits_n\left(|\Delta E_{na}|-\Delta E_{na}\right) = 2\sum\limits_{n<a}\beta_{an}$ and considering that the level width is related to the imaginary part of the radiative shift corresponding to the one-loop self-energy correction of the bound electron as
\begin{eqnarray}
\label{22}
-2\mathrm{Im}\Delta E_a = \Gamma_a,
\end{eqnarray}
one can arrive at
\begin{eqnarray}
\label{23}
\delta\Gamma_a =4e^2\sum\limits_{n<a}\Delta E_{na}^2
\left|\langle a | \vec{r} | n\rangle\right|^2\left[\langle n| \vec{\mu}\vec{B} |n \rangle\right.
\\\nonumber
\left.
-\langle a| \vec{\mu}\vec{B} |a \rangle\right].\qquad
\end{eqnarray}

\section{Basic formulas for the radiative correction to the level width in the presence of a magnetic field combined with blackbody radiation}
\label{apB}
\renewcommand{\theequation}{B\arabic{equation}}
\setcounter{equation}{0}

Next, we demonstrate another possible way to obtain a radiative correction corresponding to the Feynman diagrams shown in Fig.~\ref{fig1a} for graphs where the usual photon loop is replaced by a thermal one. In this case the photon propagator (\ref{10}) should be replaced by
\begin{eqnarray}
\label{24}
D_{\mu\nu}^{\beta}(x_{1}x_{2})=-\frac{g_{\mu\nu}}{\pi r_{12}}\int\limits_{-\infty}^{\infty}d\omega\, n_{\beta}(|\omega|)
\\\nonumber\times
\sin(|\omega|r_{12})e^{-\mathrm{i}\omega(t_{1}-t_{2})},\qquad
\end{eqnarray}
where $n_{\beta}(\omega)=(\mathrm{exp}(\beta\omega)-1)^{-1}$ is the photon density number of BBR field (Planck distribution function), $\beta=1/k_{\mathrm{B}}T$, $k_{\mathrm{B}}$ is the Boltzmann constant in relativistic units, $T$ is the radiation temperature in kelvin and $r_{12}\equiv |\vec{r}_1-\vec{r}_2|$. The details of the thermal photon propagator derivation in the form Eq. (\ref{24}) can be found in \cite{SLP-QED,S-2020}.

Insertion of the expression (\ref{24}) into Eqs. (\ref{5})-(\ref{8}) leads to the expressions \cite{g-ZGS}:
\begin{eqnarray}
\label{25}
\Delta E_{a}^{\mathrm{ver},\beta}
=-\frac{e^3 }{\pi}\sum\limits_{\pm}
\sum\limits_{n, m}\int\limits_{0}^{\infty}d\omega n_{\beta}(\omega)
\\
\nonumber
\times
\frac{\langle am| \frac{1-\vec{\alpha}_1\vec{\alpha}_2}{r_{12}} \sin(\omega r_{12}) |na\rangle
\langle n| \gamma^\nu A^{\rm pert}_\nu|m \rangle}{[E_{a}\pm\omega-E_{n}(1-\mathrm{i}0)][E_{a}\pm\omega-E_{m}(1-\mathrm{i}0)]},
\end{eqnarray}
where $\sum\limits_{\pm}$ denotes the sum of two contributions with $+$ and $-$ in energy denominators. Similar evaluation for the wave function part gives
\begin{eqnarray}
\label{26}
\Delta E_{a}^{\mathrm{wf},\beta} = -\frac{2e^3 }{\pi}\sum\limits_{\pm}  
\int\limits_{0}^{\infty} d\omega\, n_{\beta}(\omega)
\\
\nonumber
\times
\left[ 
\sum\limits_{\stackrel{n, m}{m\neq a}}\frac{\langle a n| \frac{1-\vec{\alpha}_1\vec{\alpha}_2}{r_{12}} \sin(\omega r_{12}) |n m\rangle
\langle m| \gamma^\nu A^{\rm pert}_\nu |a \rangle}{[E_{a}\pm\omega-E_{n}(1-\mathrm{i}0)][E_{a}-E_{m}]}
\right.
\\
\nonumber
\left.
- \frac{1}{2}\sum\limits_n\frac{\langle a n| \frac{1-\vec{\alpha}_1\vec{\alpha}_2}{r_{12}} \sin(\omega r_{12}) |n a\rangle
\langle a| \gamma^\nu A^{\rm pert}_\nu |a \rangle}{[E_{a}\pm\omega-E_{n}(1-\mathrm{i}0)]^2}
\right].
\end{eqnarray}

Then, again for the specific interaction $\vec{\mu}\vec{B}$, see Eqs. (\ref{13}), (\ref{14}), one can obtain
\begin{eqnarray}
\label{27}
\Delta E_a^\beta = -\frac{e^3}{\pi}\sum\limits_{\pm}\sum\limits_n \left[\langle n| \gamma^\nu A^{\rm pert}_\nu |n \rangle-\langle a| \gamma^\nu A^{\rm pert}_\nu |a \rangle\right]\qquad
\\
\nonumber
\times
\int\limits_{0}^{\infty} d\omega\, n_{\beta}(\omega)\frac{\langle a n| \frac{1-\vec{\alpha}_1\vec{\alpha}_2}{r_{12}} \sin(\omega r_{12}) |n a\rangle}{[E_{a}\pm\omega-E_{n}(1-\mathrm{i}0)]^2}.\qquad
\end{eqnarray}
Integration over the frequency $\omega$ can be performed using Eq. (\ref{16}) followed by application of the Sokhotski-Plemelj theorem, or using the generalized expression:
\begin{eqnarray}
\label{28}
\frac{1}{(x+\mathrm{i}\epsilon)^2}\xrightarrow[\epsilon\rightarrow+0]{} \mathrm{P.V.}\frac{1}{x^2} + \mathrm{i}\pi \delta'(x),
\end{eqnarray}
where $'$ in the $\delta$-function denotes the derivative with respect to $x$, and $\mathrm{P.V.}$ corresponds to the principal value.

Since the real and imaginary parts of $\Delta E_a^\beta$ are determined through the energy denominator in Eq. (\ref{27}), to obtain a correction to the level width, only the part containing the derivative of the $\delta$-function can be considered (the real part of (\ref{27}) was considered in \cite{g-ZGS}). This leads to
\begin{eqnarray}
\label{29}
\mathrm{Im}\Delta E_a^\beta = -e^2\sum\limits_n \left[\langle n| \vec{\mu}\vec{B} |n \rangle-\langle a| \vec{\mu}\vec{B} |a \rangle\right]
\\
\nonumber
\times
\frac{\partial}{\partial E_a}n_{\beta}(|\Delta E_{na}|)\langle a n| \frac{1-\vec{\alpha}_1\vec{\alpha}_2}{r_{12}} \sin(|\Delta E_{na}| r_{12}) |n a\rangle,
\end{eqnarray}
where, in opposite to Eq. (\ref{23}), the state $n$ can be lower and higher in respect to $a$.

Employing again the nonrelativistic limit for the matrix element $\langle a n|\dots| n a\rangle$, one can find
\begin{eqnarray}
\label{30}
\langle a n| \frac{1-\vec{\alpha}_1\vec{\alpha}_2}{r_{12}} \sin(|\Delta E_{na}| r_{12}) |n a\rangle
\\\nonumber
\approx -\frac{2}{3}|\Delta E_{na}|^3|\langle a|\vec{r}|n\rangle|^2.\qquad
\end{eqnarray}
Insertion of (\ref{30}) into Eq. (\ref{29}) with the use of Eq. (\ref{22}) yields
\begin{eqnarray}
\label{31}
\delta\Gamma^\beta_a = \frac{4e^2}{3}\sum\limits_n \left[\langle n| \vec{\mu}\vec{B} |n \rangle-\langle a| \vec{\mu}\vec{B} |a \rangle\right]|\langle a|\vec{r}|n\rangle|^2\qquad
\\
\nonumber
\times
\left[\beta \Delta E_{na}^3 n^2_\beta(|\Delta E_{na}|) e^{\beta\, |\Delta E_{na}|} - 3 \Delta E_{na}^2 n_\beta(|\Delta E_{na}|)\right].
\end{eqnarray}
Formula (\ref{31}) represents the main result discussed in section \ref{B_BBR}.

\end{document}